\documentclass[useAMS,usenatbib]{mn2e}

\def\aap{A\&A}
\def\apj{ApJS}

\def\apjs{ApJ Supp.}
\def\apss{Ap\&SS}
\def\mnras{MNRAS}
\def\aj{AJ}
\def\nat{Nature}

\def\pasp{PASP}
\def\pasj{PASJ}
\def\procspie{Proc.~SPIE}

\usepackage{amsmath}
\usepackage{graphicx}
\usepackage{color}
\usepackage{longtable,lscape}

\def\kms{$\mbox{km s}^{-1}$}

\newcommand\ion[2]{#1$\;${\scshape{#2}}}
\usepackage{txfonts}

\title[Discovery of a Be/X-ray pulsar binary and associated supernova remnant in the Wing of the SMC]{Discovery of a Be/X-ray pulsar binary and associated supernova remnant in the Wing of the SMC}
\author[V. H\'{e}nault-Brunet et al.]{V. H\'{e}nault-Brunet$^{1}$\thanks{E-mail: vhb@roe.ac.uk}, L.\,M. Oskinova$^{2}$, M.\,A. Guerrero$^{3}$, W. Sun$^{4}$, Y.-H. Chu$^{5}$,\newauthor C.\,J. Evans$^{6,1}$, J.\,S.~Gallagher~III$^{7}$, R.\,A. Gruendl$^{5}$, J.~Reyes-Iturbide$^{8}$\\
$^{1}$Scottish Universities Physics Alliance (SUPA), Institute for 
Astronomy, University of Edinburgh, Blackford Hill, Edinburgh, EH9 3HJ, UK\\
$^{2}$Institute for Physics and Astronomy, University of Potsdam,
14476 Potsdam, Germany\\
$^{3}$Instituto de Astrof\'\i sica de Andaluc\'\i a, IAA-CSIC. 
c/ Glorieta de la Astronom\'\i a s/n, 18008 Granada, Spain\\
$^{4}$Department of Astronomy, Nanjing University, Nanjing, 210093 Jiangsu, China\\
$^{5}$Department of Astronomy, University of Illinois, 1002 West Green Street, Urbana, IL 61801, USA\\
 $^{6}$UK Astronomy Technology Centre, 
           Royal Observatory Edinburgh, 
           Blackford Hill, Edinburgh, EH9 3HJ, UK\\
$^{7}$Department of Astronomy, University of Wisconsin-Madison,
5534 Sterling, 475 North Charter St., Madison, WI 53706, USA\\            
$^{8}$Escuela Superior de F\'\i sica y Matem\'aticas, IPN, 
    U.P.\ Adolfo L\'opez Mateos, C.P. 07738 D.F., Mexico}

\begin{document}

\date{Accepted 2011 October 27. Received 2011 October 27; in original form 2011 October 12}

\pagerange{\pageref{firstpage}--\pageref{lastpage}} \pubyear{2011}

\maketitle

\label{firstpage}

\begin{abstract}
{We report on a new Be/X-ray pulsar binary located in the Wing
  of the Small Magellanic Cloud (SMC). The strong pulsed X-ray source
was discovered with the {\em Chandra} and {\em XMM-Newton} X-ray observatories. The X-ray pulse period
  of 1062\,s is consistently determined from both
  {\em Chandra} and {\em XMM-Newton} observations, revealing one of the slowest rotating X-ray pulsars known in the SMC. The optical counterpart of the X-ray source is the emission-line star 2dFS\,3831. Its B0-0.5(III)e+ spectral type is determined from VLT-FLAMES and 2dF optical spectroscopy, establishing the system as a Be/X-ray binary (Be-XRB). The hard X-ray spectrum is well fitted by a power-law with additional thermal and blackbody components, the latter reminiscent of persistent Be-XRBs. This system is the first evidence of a recent supernova in the low density surroundings of NGC\,602. We detect a shell nebula around 2dFS\,3831 in H$\alpha$ and [\ion{O}{iii}] images and conclude that it is most likely a supernova remnant. If it is linked to the supernova explosion that created this new X-ray pulsar, its kinematic age of $(2-4)\times10^4$~yr provides a constraint on the age of the pulsar.}

\end{abstract}


\section{Introduction}

The Small Magellanic Cloud (SMC) is host to a large population of
$\sim$50 high-mass X-ray binaries (HMXBs), comparable to the number
known in the Galaxy \citep[e.g.][]{haberl, Coe:2005, Reig:2011}. However,
unlike the Galactic population, all of these systems but one are Be/X-ray binary (Be-XRB) systems, with the exception (SMC-X1) containing a supergiant
companion instead \citep{webster:1972}.

The Wing of the SMC is the region linking the eastern side of the SMC
to the Bridge. Despite its lower content of gas, dust, and stars, it
had a major star forming event $\sim$11 Myr ago and appears, as expected, deficient in Be-XRBs \citep{Antoniou:2010}. Note however that the coverage of the SMC Wing by X-ray observations has been
sparser than that of the Bar. Each new discovery of a Be-XRB in the SMC Wing is therefore particularly noteworthy \citep[e.g.][]{McGowan:2007}.

We have conducted {\em Chandra} and {\em XMM-Newton} observations of the far
eastern region of the Wing, concentrating on the star forming region
NGC\,602. The full description of these observations will be
presented in a forthcoming paper (Oskinova et al., in preparation). In
this letter, we concentrate on one bright X-ray source located near NGC\,602 in projection and coinciding with the emission-line star 2dFS\,3831
(R.A. =$1^{\rm h}27^{\rm m}46.03^{\rm s}$,
Dec. $=-73\degr32^{\prime}56.42^{\prime\prime}$, J2000.0), which is revealed as a Be-XRB. Based on the results of section \ref{timing}, throughout this paper we will refer to the X-ray source as SXP\,1062 following the nomenclature of \citet{Coe:2005} for SMC X-ray pulsars.

The stellar population of NGC\,602 (which comprises three star
clusters: NGC602\,A,\,B, and C) has been studied in detail. In addition to very old
stars of 6-8 Gyr, likely the SMC field population, the NGC\,602
clusters contains young stars of $\sim$4-5\,Myr
\citep{westerlund, Hutchings:1991,cignoni:2009}. The significant number of young
stellar objects detected in the NGC\,602A cluster
\citep{carlson:2007, gouliermis:2007} suggests that star
formation is still ongoing. As for what initially triggered the
formation of these clusters, the low density environment in the Wing
of the SMC hints at an additional mechanism acting along with gravitational
collapse. NGC\,602 is located at the intersection of three \ion{H}{i}
shells, and \citet{Nigra:2008} suggested that its formation is the
result of the interaction of two expanding shells $\sim$7 Myr
ago. \citet{Schmalzl:2008} proposed instead that star
formation was possibly induced by encounters with the Large
Magellanic Cloud or the Milky Way. NGC\,602 is associated with SGS-SMC1, the only H$\alpha$ supergiant shell known in the SMC. The discovery of a Be-XRB
in this area (only $\sim$7$^{\prime}$ to the west of NGC\,602A) provides important clues to the star formation history in
the Wing and the evolution of large-scale structures in the
interstellar medium.

This letter is organized as follows. We describe the X-ray
observations and analysis in section \ref{x-ray}. In section
\ref{optical}, we present the spectroscopy of the optical counterpart. We
discuss the properties and environment of SXP\,1062 in section \ref{discussion} and present
our conclusions in section \ref{conc}.

\section{X-ray observations} \label{x-ray}

The X-ray data were obtained with the ACIS-I camera on the {\em Chandra}
X-ray Observatory and with the EPIC cameras onboard {\em
 XMM-Newton}. The {\em Chandra} observations consisted of 11 separate
exposures acquired between 2010-03-31 and 2010-04-29 (effective 
exposure time of 290.7\,ks), while 4 separate exposures were obtained 
with {\em XMM-Newton} between 2010-03-25 and 2010-04-12 (EPIC-pn
effective exposure time of 162.5\,ks).  The data were reduced using the
most up-to-date versions of the corresponding data reduction
software. The X-ray source CXO\,J012745.97-733256.5 (=SXP\,1062) coinciding with the 
optical emission-line star 2dFS\,3831 is detected in each of these 15
exposures. In the {\em Chandra} observations, the positional uncertainty
(1\,$\sigma$) of the X-ray source is $0.9''$. The source was also
detected during an \emph{XMM-Newton} slew on 2009-11-16. 
 Interestingly, the source was not detected in the {\em ROSAT All Sky
   Survey}, although at its present luminosity it would have been
 bright enough at an expected \emph{ROSAT} PSPC count rate of 0.01\,
 counts~s$^{-1}$.

\subsection{Spectral analysis} \label{xspec}

\begin{figure}
  \centering \includegraphics[width=8cm]{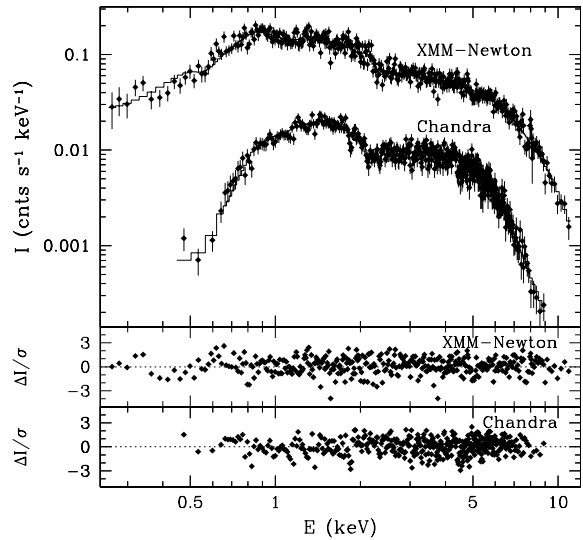}
\caption{
{\em Chandra} ACIS-I and {\em XMM-Newton} EPIC-pn spectra of SXP\,1062.
Overplotted is the best-fit model including power-law, thermal, and 
blackbody components.  
The lower panels show the residuals of the fit.  
}  
        \label{bestfit_spec}
  \end{figure}

Figure\,\ref{bestfit_spec} shows the combined background-subtracted
\emph{Chandra} and \emph{XMM-Newton} spectra of SXP\,1062.  The
\emph{Chandra} spectra were extracted from a 12\farcs8 circular region
and the background was extracted from a concentric annular region of
radii 12\farcs8 and 19\farcs2 which was free of sources.  Likewise,
the \emph{XMM-Newton} spectra were extracted from a
32\arcsec\ circular region and the background from 4 adjacent circular
regions free of sources with radius 45\arcsec.

\begin{table*}
\caption{
Best-fit parameters of the spectral models. The code for the different components of each model is: ``P'' for power-law, ``B'' for blackbody, and ``T'' for optically thin thermal emission with SMC abundances.
}
\label{tab:fit}
\medskip
\begin{tabular}{lcccccccc} \hline\hline

Model  &     $N_{\rm H}$       &    $\Gamma$     &         $A_{\rm power law}$          &     
$kT$      &       $A_{\rm blackbody}$                  &      $kT$       
&           $A_{\rm thermal}$      &    $\chi^2$/DoF   \\
     & (10$^{21}$ cm$^{-2}$) &                 &(phot~keV$^{-1}$~cm$^{-2}$~s$^{-1}$)  &     
(keV)     &(10$^{39}$ erg~s$^{-1}$~kpc$^{-2}$)    &      (keV)      
&      (10$^{9}$ cm$^{-5}$)       &                   \\
\hline
P     &    1.18$\pm$0.06     & 0.746$\pm$0.011 &  (6.60$\pm$0.11)$\times$10$^{-5}$   &    
$\dots$    & 	    $\dots$		&	      $\dots$    
&		   $\dots$	    & 1.30 \\  
                                                                   
PB    &    1.48$\pm$0.12     & 1.185$\pm$0.022 &  (6.35$\pm$0.35)$\times$10$^{-5}$  & 
2.13$\pm$0.05 & (1.11$\pm$0.20)$\times$10$^{-5}$ &
$\dots$     &  	    $\dots$	       & 1.21  \\

PT    &    1.63$\pm$0.04     & 0.732$\pm$0.011 &  (6.48$\pm$0.12)$\times$10$^{-5}$  &    
$\dots$    & 	    $\dots$			  & 0.589$\pm$0.032 & 
2.60$\pm$0.32 & 1.19 \\

PBT   &    1.35$\pm$0.12     & 0.767$\pm$0.027 &  (4.87$\pm$0.30)$\times$10$^{-5}$  & 
1.54$\pm$0.16 &  (4.5$\pm$0.9)$\times$10$^{-6}$  & 
0.648$\pm$0.029 & 2.80$\pm$0.30 & 1.07 \\

\hline\hline

\end{tabular}
\end{table*}

As a first step the background-subtracted \emph{Chandra} and
\emph{XMM-Newton} spectra of SXP\,1062 were simultaneously fitted
using a simple absorbed power-law model. The best-fit parameters, listed in Table\,\ref{tab:fit}, imply an 
absorbed flux in the energy range 0.2-12.0 keV of $f_{\rm
 X}$=1.8$\times$10$^{-12}$~erg~cm$^{-2}$~s$^{-1}$.  This corresponds
to an intrinsic X-ray luminosity in this same energy range of $L_{\rm
 X}$=6.9$\times$10$^{35}$~erg~s$^{-1}$ assuming a distance modulus
of 18.7 appropriate for the Wing \citep[e.g.][]{cignoni:2009}, or
$L_{\rm X}$=8.2$\times$10$^{35}$~erg~s$^{-1}$ for the `standard' SMC
distance modulus of 18.9 \citep[e.g.][]{harries:2003}.  
The photon index $\Gamma$ of $\sim$0.75 is typical of X-ray pulsar 
binaries \citep[e.g.][]{yokogawa:2003} and a signature of accretion onto 
a strongly magnetized neutron star.

The simple power-law model suggests some emission excess below 1 keV and at high energies.  As a next step, we added a blackbody component, which statistically
improved the fit (see Table\,\ref{tab:fit}). The
observed X-ray flux is similar to that of the previous model, but the
X-ray luminosity decreases by $\sim10\%$. We also tried a model for which a
thermal plasma component with SMC abundances is included instead of a
blackbody. Similarly to the blackbody, the addition of a thermal
component to the power-law model improves the fit. In this case, the
intrinsic X-ray luminosity is only marginally greater by $\sim1\%$
compared to the simple power-law model.

Finally, the best fit is achieved by adding simultaneously a thermal and 
a blackbody component to the power-law (see Table\,\ref{tab:fit} and Figure\,\ref{bestfit_spec}).  
The flux for this model (PBT in Table \ref{tab:fit}) is $f_{\rm X}$=1.66$^{+0.19}_{-0.25}\times$10$^{-12}$
erg~cm$^{-2}$~s$^{-1}$, corresponding to an intrinsic luminosity $L_{\rm
 X}=$6.3$^{+0.7}_{-0.8}\times$10$^{35}$~erg~s$^{-1}$ for a distance modulus
of 18.7.

The column densities are well constrained by spectral fitting, albeit 
the values slightly differ depending on the model. 
Using the best-fit values, we note that the H column densities 
in Table\,\ref{tab:fit} imply 
$A_V=0.63-0.87$ mag for the $N_{\rm H}$/$A_V$ ratio of 
$1.87\times10^{21}$\,cm$^{-2}$, i.e. $E_{B-V}=0.20-0.28$ 
assuming $R_{V}=3.1$.  
This range is in good agreement with the $E_{B-V}=0.19$ value 
computed by comparing the value $(B-V)=-0.04$ of 2dFS\,3831 \citep{massey:2002} with 
the intrinsic value $(B-V)_0=-0.23$ of a B0-0.5III star \citep{wegner:1994} given 
the uncertainties on the photometry and spectral type (see section \ref{optical} for the optical spectroscopy of 2dFS\,3831).

\subsection{Timing Analysis} \label{timing}

The photon arrival times were corrected for the solar system
barycenter.  We searched for pulsations in the X-ray light curves in
the soft and hard energy bands (0.4\,keV---1.5\,keV,
2.0\,keV---7.2\,keV) and a total energy band (0.4\,keV---7.2\,keV),
using fast Fourier transform and light-curve folding techniques as
implemented in the timing analysis software {\em xronos}.

\begin{figure}
   \centering
   \includegraphics[width=8cm]{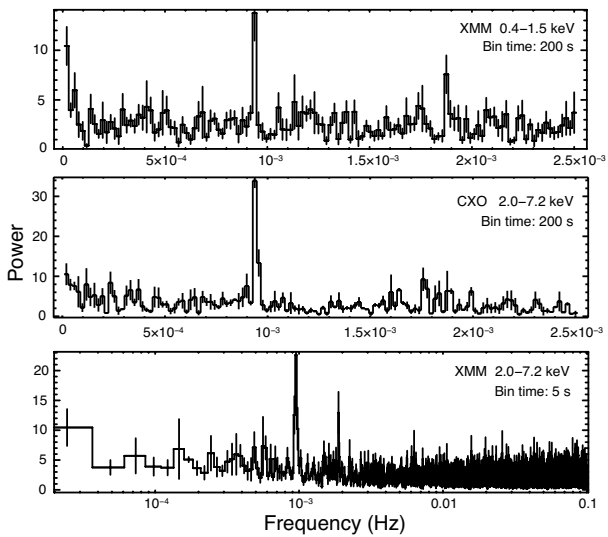}
      \caption{{\em XMM-Newton} and {\em Chandra} power density spectra in the soft and hard energy bands for different bin times.}          \label{power}
   \end{figure}

\begin{figure}
      \centering
   \includegraphics[width=8cm]{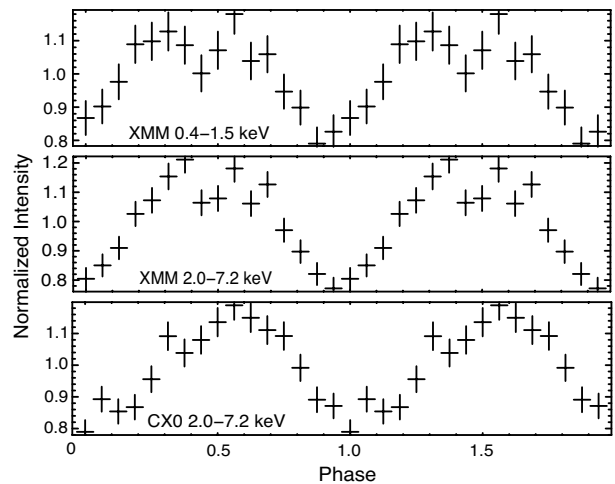}
      \caption{ Pulse profiles folded with a period of 1062\,s in different energy bands.}
      \label{pulse}
   \end{figure}

Figure\,\ref{power} shows the inferred {\em Chandra} and {\em
  XMM-Newton} power density spectra in the soft and hard energy bands
with a clear peak at a frequency of $9.4\times 10^{-4}$\,Hz (P=1062\,s). This coherent X-ray pulse period establishes the
source as a binary X-ray pulsar. The pulse profiles folded with this period are shown
in Figure \ref{pulse}.

The light-curves used to compute the  {\em XMM-Newton} power spectrum in
the hard energy band, the {\em XMM-Newton} power spectrum in the soft energy band, and the {\em Chandra} power spectrum in the hard energy band were binned by respectively 5\,s, 200\,s, and 200\,s.
The 5\,s bin time for one of these was chosen to make sure that a
shorter pulse was not missed. No pulse was found in the soft band {\em
  Chandra} observations, possibly because the count rate is too
low. Because of the softer response of the EPIC-pn camera, the pulse
in the soft band is the most obvious in the light-curves obtained with
this instrument.

Note that in addition to the pulse period, X-ray variability was also
detected at a level of $\sim20\%$ peak-to-peak on a timescale of
several days in both {\em XMM-Newton} and {\em Chandra}
light-curves. Given that this variability does not appear regular and that
the time sampling is very sparse, we do not analyse these variations
further.

\begin{figure*}
   \centering
   \includegraphics[width=16.7cm]{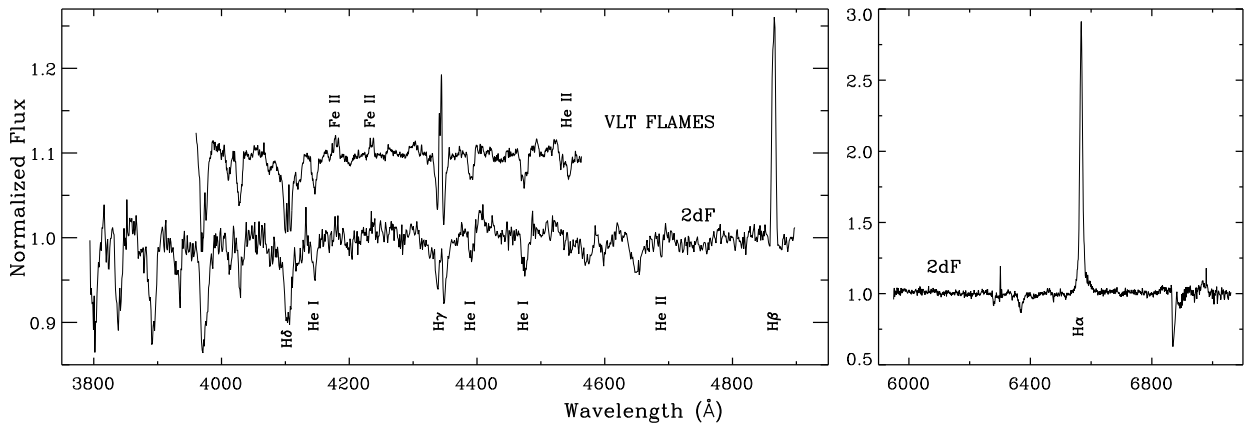}
      \caption{Optical VLT FLAMES and 2dF spectra of 2dFS\,3831 with key 
spectral features identified. The VLT FLAMES spectrum is offset vertically.}
         \label{opt}
\end{figure*}

\section{Optical spectroscopy} \label{optical}

From its X-ray position with subarcsecond precision from {\em Chandra}
images, the optical counterpart to SXP\,1062 is
identified as 2dFS\,3831 \citep{evans:2004}. We observed this star
with the VLT FLAMES instrument \citep{pasquini:2002} on 2010 October
25 as part of a spectroscopic survey of massive stars in NGC\,602
complementing our X-ray observations. Spectra were obtained in the
Medusa-fibre mode of FLAMES using the LR02 setting of the Giraffe
spectrograph \citep[3960--4564 \AA, R=7000, e.g.][]{evans:2011}. Five pairs of 1800\,s exposures were
obtained. The standard data processing (bias subtraction, fibre
location, summed extractions, division by flat-field, wavelength
calibration) was done using the ESO Common Pipeline Library FLAMES
reduction routines (v.2.8.7). Additionally, heliocentric correction
and subtraction of a median sky spectrum were performed \citep[for
  details see e.g.][]{evans:2011}. The spectra from all the exposures
were then normalized and merged. The 2dF spectra of this star were
also retrieved. These two spectra cover the regions from $\sim$3800 to
4900 \AA\ and $\sim$6000 to 7000 \AA, with resolving powers of 1500
and 2500 respectively \citep{evans:2004}. Figure \ref{opt} shows the
VLT FLAMES and 2dF spectra of 2dFS\,3831. The VLT FLAMES spectrum has been smoothed and rebinned to an equivalent resolving power of
R=4000.

A weak \ion{He}{ii} $\lambda$4542 absorption line is visible in the
VLT FLAMES spectrum and there is a hint of a weak \ion{He}{ii}
$\lambda$4686 line in the 2dF spectrum, but \ion{He}{ii} $\lambda$4200
is absent, suggesting a spectral type around B0-0.5 following the
classification adopted by \citet{evans:2004}.

Several characteristics attributable to a circumstellar disc indicate
that 2dFS\,3831 is a classical Be star. \ion{Fe}{ii} $\lambda$4179 and
\ion{Fe}{ii} $\lambda$4233 emission lines typical of some early Be
stars \citep[e.g.][]{slettebak:1982} are visible in the VLT FLAMES
spectrum. The H$\alpha$ emission seen in the 2dF spectrum (see the right panel in Fig.\,\ref{opt}) is
relatively strong with an equivalent width (EW) of -23\,\AA. There is
significant H$\beta$ emission, plus emission in the core of the other
Balmer lines (double-peaked in the VLT FLAMES spectrum) and apparent
infilling of the \ion{He}{i} absorption lines. Two Micron All Sky
Survey \citep[2MASS;][]{skrutskie:2006} JHK$_{\rm s}$ photometry also
indicates a clear infrared excess when compared with expected colors
for a B0-0.5 spectral type \citep{wegner:1994}. The double-peaked
emission in the Balmer lines is not due to oversubtraction of nebular
features present in the median sky spectrum. Also, given the relative
weakness of the nebular emission lines in the mean sky spectrum, the
emission in the core of the Balmer lines is most likely dominated by
circumstellar material and not by nebular contamination.

The observed magnitude \citep[V=14.36, B=14.32;][]{massey:2002} and a range of reasonable
extinction estimates (including both the interstellar and
circumstellar components) lead to an absolute magnitude consistent
with a B0-0.5 giant \citep{walborn:1972, vacca:1996}. We thus determine
the spectral type of 2dFS\,3831 as B0-0.5(III)e+, where `+' signifies the presence of \ion{Fe}{ii}.
This corresponds to a typical effective temperature $T_{\rm eff}\sim26\,000$\,K and an
evolutionary mass $M\sim15\, {\rm M}_{\odot}$ at SMC metallicity
\citep[cf.][]{trundle:2007}.

We estimated the radial velocity of 2dFS\,3831 from the VLT FLAMES
spectrum by fitting Gaussian profiles to the wings of H$\gamma$,
H$\delta$, H$\epsilon$, \ion{He}{i} $\lambda$4143, and \ion{He}{i}
$\lambda$4388. The average of all measurements is 167 \kms with a
standard deviation of 13 \kms. This is consistent with the
  mean velocity of massive stars in the SMC \citep{evans:2008}, and
  also with the gas velocities measured by \citet{Nigra:2008} across
  the N\,90 \ion{H}{ii} region ionized by NGC\,602A. Thus, we do not see any evidence that the HMXB acquired a high space velocity following the supernova explosion.

\section{Discussion }\label{discussion}

\subsection{The properties of SXP\,1062}

SXP\,1062 is only
the third SMC X-ray pulsar with a spin period larger than 1000\,s \citep{laycock:2010}. These slowly rotating pulsars are
particularly interesting because they represent a challenge for the
theory of spin evolution of a neutron star in a close binary system
\citep[e.g.][]{ikhsanov:2007}.

According to the Corbet diagram for SMC Be-XRBs relating spin and
orbital period \citep{corbet:2009}, we expect SXP\,1062 to have a
binary period of $\sim$300 days. Such a long orbital period is also
expected from the $P_{\rm orb}$-EW(H$\alpha$) diagram
\citep{reig:1997}, from which our measured EW(H$\alpha$) suggests an
orbital period of $\sim$100 days. Note however that the maximum
EW(H$\alpha$) of the system, which probes the maximum size of the disc
(and indirectly the orbital period), could be higher than our
instantaneous measurement, so this period estimate should be taken as
a lower limit. 

SXP\,1062 shares many characteristics with the class of persistent
Be-XRBs \citep[e.g.][]{Reig:2011}: a relatively low X-ray luminosity
$\sim10^{34}-10^{35}$\,erg\,s$^{-1}$, a slowly rotating pulsar with
$P>200$\,s, a relatively flat light curve with sporadic and unpredicted increases in intensity by less than an order of magnitude, and a lack of iron lines at $\sim$6.4\,keV indicative of small amounts of material in the vicinity of the neutron star. A thermal excess of
blackbody type, with a high temperature ($kT >1$\,keV) and a small
emission area ($R<0.5$\,km, consistent with a hot spot at the polar cap of the neutron star) has recently been suggested as a
common feature of persistent Be-XRBs \citep{lapalombara2009}. Such a component is also identified in the X-ray spectrum of SXP\,1062, for which we infer a size of $\sim0.4$\,km for the
blackbody source from our best-fit models (section\,\ref{xspec}).

\subsection{The environment of SXP\,1062}

\begin{figure}
  \centering
  \includegraphics[width=8cm]{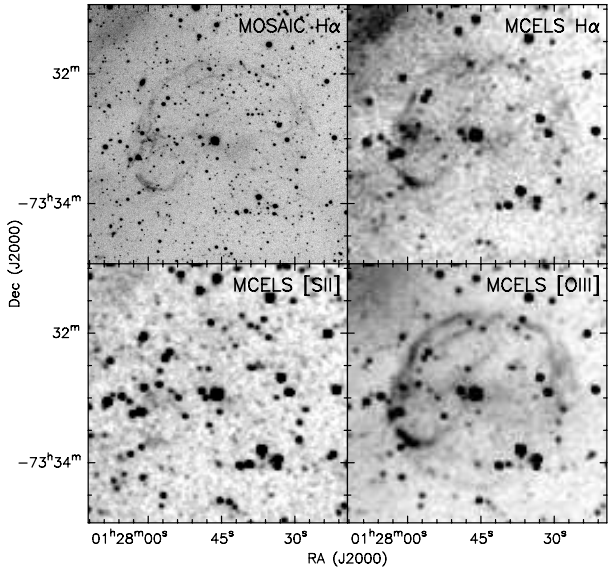}
  \caption{MOSAIC H$\alpha$ image and MCELS H$\alpha$, [\ion{S}{ii}], and [\ion{O}{iii}] images of a region centered on the position of 2dFS\,3831=SXP\,1062 and showing the shell nebula detected around the target.}
        \label{MCELS}
\end{figure}

A shell nebula is detected around SXP\,1062 in the Magellanic Cloud Emission-line Survey \citep[MCELS;][]{Setal99} H$\alpha$ image and in the higher resolution CTIO 4m MOSAIC \citep{Metal98} H$\alpha$ image (see Figure~\ref{MCELS}). This shell is also detected in the MCELS [\ion{O}{iii}] image, but hardly in [\ion{S}{ii}]. We adopt below a distance modulus of 18.7. The shell is not uniform, with radii varying from 75$''$ at
the sharp rim on the northeast to 94$''$ at the diffuse edge on the southwest.
The peak surface brightness of the northeast rim is
$\sim 6 \times10^{-17}$ ergs cm$^{-2}$ s$^{-1}$ arcsec$^{-2}$, corresponding to an emission measure of 30 cm$^{-6}$ pc.  The width of the rim suggests that
the shell thickness ($\Delta R$) is 5--10\% of the shell radius ($R$).
The longest emitting length at the shell rim is $2R[1-(\Delta R/R)^2]^{1/2}$; thus, the rms density in the shell is 1.3$\pm$0.3 H-atom cm$^{-3}$.  The gas
mass in the shell is 250$\pm$100 $M_\odot$.  This large mass indicates that
the shell gas must be dominated by interstellar material.

The shell morphology resembles supernova remnants (SNRs) in the Magellanic Clouds. The X-ray images also suggest that diffuse X-ray emission possibly associated with a SNR may be present in the vicinity of SXP\,1062 (Oskinova et al., in preparation). As the shell mass is much larger than the typical supernova ejecta
mass, we assume that the shell is a SNR in the Sedov phase.  The kinetic
energy in the shell would be $\sim$30\% of the supernova explosion energy.
Adopting a canonical explosion energy of 10$^{51}$ ergs, the current shell
kinetic energy is $3\times10^{50}$ ergs, and implies a shell expansion velocity
of 350$\pm$100 km~s$^{-1}$ and an age of 0.4($R/V$)=$(2-4)\times10^4$~yr. This age is much larger than the cooling timescale of this low density
gas, justifying an adiabatic shock for the Sedov phase.  The pre-shock
interstellar gas density, 1/4 the shell density, is $0.3\pm0.1$ H-atom cm$^{-3}$, consistent with the low density expected in
the SMC Wing.  The bright [\ion{O}{iii}] emission can be easily produced
by a 350 km~s$^{-1}$ shock \citep[cf.][]{Hetal87}. The ionizing flux of the B0--0.5III star can easily photoionize the shell gas and its surrounding pre-shock medium; furthermore, the diffuse [\ion{O}{iii}] emission to the northeast exterior of the shell indicates the existence of a harsh radiation field; therefore,
the ionization stage of the pre-shock medium may be too high to produce
strong post-shock [\ion{S}{ii}] emission. We thus conclude that the shell nebula detected around SXP\,1062 is most likely a SNR. The extent of the shell, which has not reached the NGC\,602 region, and its young kinematic age suggest that this supernova event did not trigger the formation of NGC\,602.

\section{Conclusion }\label{conc}

We reported the discovery of a Be-XRB, SXP\,1062, containing one of the slowest rotating X-ray pulsars in the SMC. We 
suggest that the shell nebula detected around this object is a SNR. Its estimated kinematic age of $(2-4)\times10^4$~yr  is probably the age of the pulsar. This is, to our knowledge, the first discovery of a pulsar associated with a supernova remnant in the SMC.

\section*{Acknowledgments}

 This study is based on observations obtained with {\em XMM-Newton},
 an ESA science mission with instruments and contributions directly
 funded by ESA Member States and NASA, and the {\em Chandra} science
 mission.  The software provided by the {\em Chandra} X-ray Center
 (CXC) in the application package CIAO and by the {\em XMM-Newton} in
 the package SAS was used. This research has made use of NASA's
 Astrophysics Data System Service and the SIMBAD database, operated at
 CDS, Strasbourg, France. We thank Frank Winkler for providing the SMC MCELS images. VHB acknowledges support from SUPA and NSERC. LMO. acknowledges support
 from the DLR grant 50\,OR\,0804. MAG is supported by the MICINN grant AYA2008-01934 that includes FEDER funds. W.S. acknowledges support
 from the DAAD grant A/10/95420. The DFG grant OS\,292/3-1 supported the project workshop. YHC, JSG, and RAG acknowledge the support of NASA grant SAO GO0-11025X.

\label{lastpage}

\end{document}